# A Search for Alternative Solid Rocket Propellant Oxidizers

Pujan Biswas[1], Parmanand Ahirwar[2], S. Nandagopal[4], Arvind Kumar[4], I. N. N. Namboothiri[3], Arindrajit Chowdhury[2], Neeraj Kumbhakarna[2]

[1] Department of Aerospace Engineering, IIT Bombay

[2] Department of Mechanical Engineering, IIT Bombay

[3] Department of Chemistry, IIT Bombay

Powai, Mumbai, Maharashtra, 400076, India

[4] High Energy Materials Research Laboratory, DRDO

Pashan, Pune, Maharashtra, 411021, India

## Abstract

Carbon-based caged and heterocyclic compounds tend to have strained molecular structures leading to high heats of formation and energetic behavior. In the current paper, molecular modelling calculations for 10 caged compounds of this type along with 2 strained aliphatic compounds and 4 simple cyclic chains are presented in view of their possible use as oxidizers in propulsion applications. Density functional theory (B3LYP) was employed for the geometry optimization of the proposed molecular structure using the 6-311++G(d,p) basis set. Heats of formation of the compounds were calculated using the molecular modeling results and their specific impulses were computed using the NASA CEA [1] software package to evaluate their potentials as propellant oxidizers.

## 1 Introduction

Ammonium perchlorate (AP) has been used extensively as an oxidizer in various solid propulsion systems. However, as Trache et al. [2] mention in their review, oxidizers that can substitute AP are being researched upon. This is further emphasized by Kettner et al. [3]. They have discussed the molecules containing trinitromethyl groups as possible oxidizers due to their ability to impart higher oxygen balance. Yu et al. [4] proposed novel energetic fluoride-based oxidizers with higher specific impulses. Although, some progress has been made, there is a wide range of possible molecules that can potentially be considered as oxidizers. Therefore, our current work examines some novel molecules in an endeavor to find a candidate than can possibly be better than AP in environmental as well as propulsive performance.

The objective of our work is to carry out a preliminary analysis of various envisioned molecular structures such that some of them can be shortlisted for synthesis and subsequent scale up. Our previous work [5] on high energy density materials (HEDMs) as possible fuels highlighted the use of strained carbon structures. The conclusions derived from the computation emphasized on the use of cubane and bis-homocubane derivatives as plausible fuels. The current work has been carried out to load oxygen-rich groups on moderately strained structures. A large number of molecules based on a wide range of structures were ideated and modelled. The compounds were evaluated based on propulsive properties such as characteristic velocity ($C^*$) and specific impulse ($I_{sp}$). The current work lists structures of 5 classes of novel compounds, among the ones studied, with promising characteristics. The classes are (i) glycouril derivatives, (ii) maleic anhydride derivatives, (iii) bis-homocubane derivatives, (iv) simple cyclic and heterocyclic and (v) aliphatic structures. The optimized structures of these compounds have been reported in Table 1. Plausible explanations for the certain behavior of these compounds have been provided.

## 2 Computational method

Molecular structures were optimized using quantum mechanics based calculations by employing Gaussian 09 [6]. We have employed B3LYP/6-311++G(d,p) level of theory consisting of triple split valence basis set with additional polarized functions. Our choice of the aforementioned level of theory is to maintain a balance between computational time and accuracy. Propulsive properties were calculated using NASA CEA on the basis of heats calculated using Gaussian. The detailed procedure of all calculations is described in our previous work [5].

| Sr No. | Structure | Name |
|---|---|---|
| | Glycouril derivatives | |
| 1. | 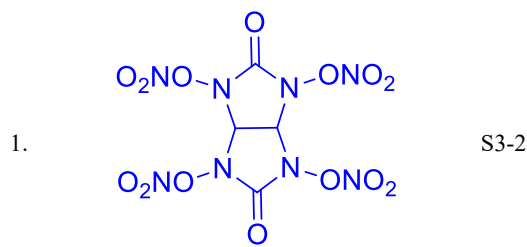 | S3-2 |

Neeraj Kumbhakarna Fax: +91-22-2572-6875
E-mail address: neeraj[underscore]k@iitb.ac.in

| | | |
|---|---|---|
| 2. | 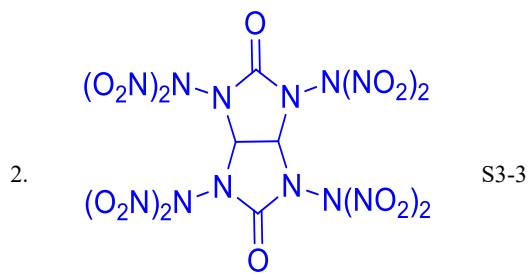 | S3-3 |
| 3. | 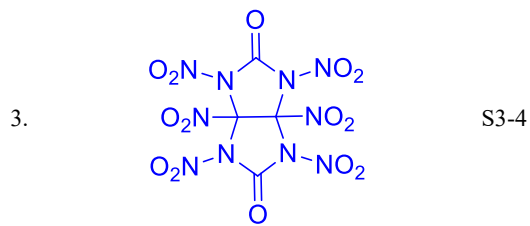 | S3-4 |
| 4. | 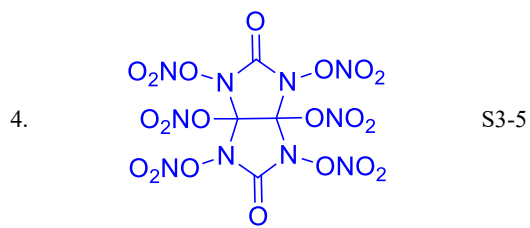 | S3-5 |
| 5. | 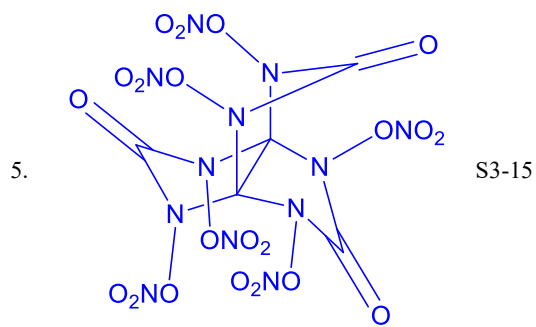 | S3-15 |

| Maleic anhydride derivatives | | |
|---|---|---|
| 6. | 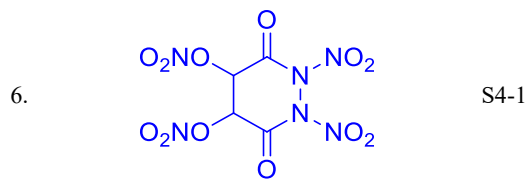 | S4-1 |
| 7. | 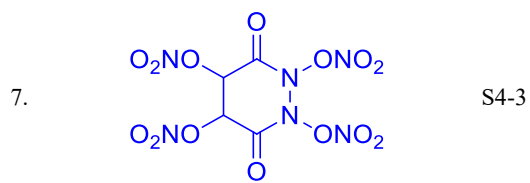 | S4-3 |
| 8. | 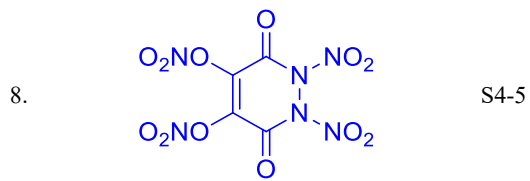 | S4-5 |
| 9. | 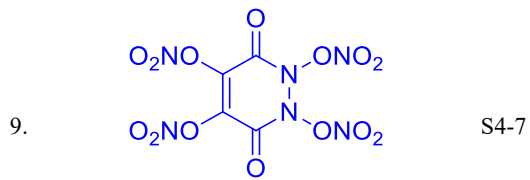 | S4-7 |

| Aliphatic structures | | |
|---|---|---|
| 10. | 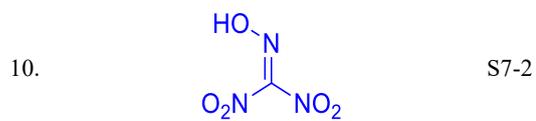 | S7-2 |
| 11. | 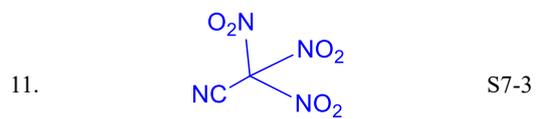 | S7-3 |

| Simple cyclic and heterocyclic | | |
|---|---|---|
| 12. | 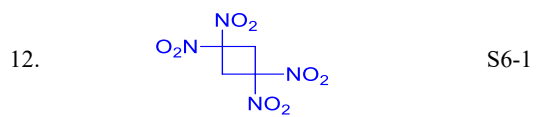 | S6-1 |
| 13. | 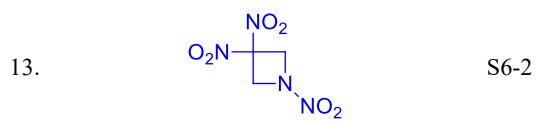 | S6-2 |
| 14. | 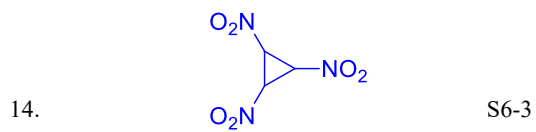 | S6-3 |
| 15. | 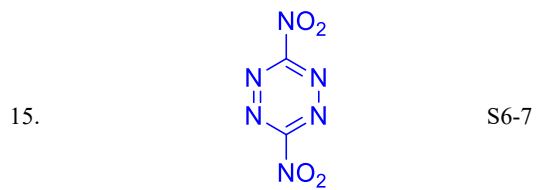 | S6-7 |

| 1,3-bishomocubane derivative | | |
| 16. | 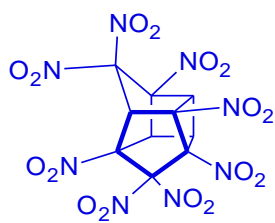 | P101 |

Table 1: Molecular structures of promising compounds

## 3 Results and Discussion

The compounds listed in Table 1 were optimized and their thermodynamic and propulsive properties were calculated. The specific impulse of these compounds were estimated by considering them to be mixed with HTPB (11% by wt.) as binder and Al (18 % by wt.) particles as fuel. Each of the oxidizers occupied 68% by wt. of the total propellant mixture. Dioctyl adipate (DOA) was considered as the plasticizer contributing to 3% by wt. The chamber pressure and temperature were set to 68.9 bar (1000 psi) and 3800K respectively during the simulations. Pressure ratio was set to 68 for the calculations. Relevant data for AP is tabulated in Table 2 for comparison with the respective properties of the analyzed novel compounds.

| A | B | C | D | E | F |
|---|---|---|---|---|---|
| AP | -295.98 | 1582.9 | 34 | 1.95 | 264.5 |

A = Compound, B = HoF (kJ/mol), C = Characteristic velocity (m/s), D = OB (%), E = Density (gm/cm$^3$), F = Specific impulse (s)

Table 2: Thermodynamic and propulsive properties of AP

### 3.1 Glycouril derivatives

Glycouril is a monomeric unit of cucurbutiril which was analyzed due to its compact structure and presence of nitrogen and oxygen atoms in the ring. The calculated data of the optimized structures are suggestive of high positive enthalpies of formation. The glycouril derivatives mentioned in Table 1 are either planar substituted glycouril structures or tricyclic structures. It was noted that S3-5 and S3-15 produce the highest specific impulses of the group. The larger number of nitrato groups along with steric hindrance in S3-15 can be a plausible explanation of this observation.

| A | B | C | D | E | F |
|---|---|---|---|---|---|
| S3-2 | 254.09 | 1644.5 | 20.73 | 1.88 | 272.8 |
| S3-3 | 1073.95 | 1668.9 | 25.62 | 1.99 | 276.6 |
| S3-4 | 449.28 | 1612.5 | 23.13 | 2.16 | 266.8 |
| S3-5 | 266.06 | 1659.5 | 37.8 | 2.01 | 278.8 |
| S3-15 | 456.27 | 1658.2 | 31.21 | 2.07 | 278.2 |

A = Compound, B = HoF (kJ/mol), C = Characteristic velocity (m/s), D = OB (%), E = Density (gm/cm$^3$), F = Specific impulse (s)

Table 3: Thermodynamic and propulsive properties of the glycouril derivatives

### 3.2 Maleic anhydride derivatives

Analysis of all the compounds in this class are indicative of good propulsive performance. However, only the ones which have their calculated specific impulses higher than AP have been reported. Steric strain produced on a single ring structure due to substituted nitro and nitrato groups can be attributed to the improved specific impulses. S4-3 and S4-7 they have the highest specific impulses in this group.

| A | B | C | D | E | F |
|---|---|---|---|---|---|
| S4-1 | 74.94 | 1601.5 | 14.72 | 1.95 | 265.8 |
| S4-3 | 77.28 | 1658.8 | 22.35 | 1.75 | 277.5 |
| S4-5 | 202.46 | 1611.9 | 19.75 | 2.12 | 266.7 |
| S4-7 | 174.77 | 1652.8 | 26.97 | 1.98 | 277.1 |

A = Compound, B = HoF (kJ/mol), C = Characteristic velocity (m/s), D = OB (%), E = Density (gm/cm$^3$), F = Specific impulse (s)

Table 4: Thermodynamic and propulsive properties of the maleic anhydride derivatives

### 3.3 Aliphatic structures

The compounds mentioned in this group are not caged structures. However, this class of compounds was chosen to determine the behavior of substituted groups which contain nitrogen and oxygen atoms independent of the ring structure. Molecules with high oxygen balance were thus ideated and scrutinized. The results are suggestive of the fact that such compounds also have potential to perform better than AP. S7-3 has the highest specific impulse amongst all the compounds presented in this paper which can be credited to the presence of three nitro groups on a single carbon. This emphasizes the importance of a compound with higher oxygen balance along with steric hindrance as a potential oxidizer.

| A | B | C | D | E |
|---|---|---|---|---|
| S7-2 | -104.01 | 1613.7 | 29.63 | 268.2 |
| S7-3 | 46.36 | 1519.7 | 18.18 | 279.9 |

A = Compound, B = HoF (kJ/mol), C = Characteristic velocity (m/s), D = OB (%), E = Specific impulse (s)

Table 5: Thermodynamic and propulsive properties of aliphatic structures

### 3.4 Simple cyclic and heterocyclic compounds

The compounds in this series are envisioned by substituting nitro groups in simple but strained cyclic chains. Data displayed in Table 6 shows that these molecules have good performance. However, it is important to note that S6-2 has higher specific impulse than S6-1 and S6-7 despite having a negative heat of formation. S6-3 is a three-carbon ring with nitro groups leading to increased strain and oxygen balance of the compound. This could be a contributing factor in having the highest specific impulse of S6-3 in this group.

| A | B | C | D | E |
|---|---|---|---|---|
| S6-1 | 310.70 | 1567.5 | -13.56 | 266.1 |
| S6-2 | -211.12 | 1436.6 | -16.67 | 267.5 |
| S6-3 | 313.17 | 1586.9 | -13.56 | 269.5 |
| S6-7 | 797.89 | 1570 | 0 | 266.2 |

A = Compound, B = HoF (kJ/mol), C = Characteristic velocity (m/s), D = OB (%), E = Specific impulse (s)

Table 6: Thermodynamic and propulsive properties of simple cyclic and heterocyclic molecules

### 3.5 1,3-bishomocubane derivative

Bishomocubane derivatives (BHCD) show promising results for their selection as propellant oxidizers. As these compounds are highly strained cage molecules that have a significant number of positions available for substitution, such characteristic can be expected. P101 is a nitro substituted BHCD but as indicated from the results in sections 3.1 and 3.2, nitrato substituted BHCD's can also prove to be potential oxidizer candidates.

| A | B | C | D | F |
|---|---|---|---|---|
| P101 | -221.42 | 1659.3 | -19.51 | 275.6 |

A = Compound, B = HoF (kJ/mol), C = Characteristic velocity (m/s), D = OB (%), E = Density (gm/cm$^3$), F = Specific impulse (s)

Table 7: Thermodynamic and propulsive properties of 1,3-bishomocubane derivative

The list of compounds presented above is representative of the extensive search that we carried out. We are also in the process of carrying out lab scale synthesis of these compounds with limited success till date.

## 4 Conclusions

The intent of the ideation, computation and analysis work presented here is to identify new series of compounds as potential oxidizers in propulsion applications. Quantum mechanics based computations to calculate the heats of formation of the proposed molecular structures were undertaken. Resulting enthalpies were further used to calculate propulsive parameters i.e. characteristic velocity and specific impulse. Our conclusions can be summarized as follows:

1. Substitution of nitro and nitrato groups substantially increase the propulsive performance of compounds.
2. Direct correlation between high enthalpies of formation and specific impulses cannot be obtained.
3. Cage compounds have the potential to perform better than ammonium perchlorate as propellant oxidizers.
4. Aliphatic and simple cyclic chains also indicate similar performance due to the high oxygen balance.
5. Calculation results presented in this work look promising. However, ease of chemical synthesis, cost, stability and compatibility with other propellant ingredients are other factors that will have to be considered.

## 5 Acknowledgment

The authors acknowledge the financial support provided by the Defence Research and Development Organisation (DRDO) and the Industrial Research and Consultancy Centre (IRCC) at the Indian Institute of Technology Bombay for this work.